\documentclass[proof]{pasj00}

\Received{$\langle$reception date$\rangle$}
\Accepted{$\langle$acception date$\rangle$}
\Published{$\langle$publication date$\rangle$}
\SetRunningHead{柱に付ける著者名}{Usage of \texttt{pasj00.cls}}

\usepackage{times}

\begin{document}

\title{Discovery of a nearby early-phase major cluster merger CIZA J1358.9-4750 }
\author{Yuichi \textsc{Kato},\altaffilmark{1}
             Kazuhiro \textsc{Nakazawa},\altaffilmark{1}
             Liyi \textsc{Gu},\altaffilmark{2}
             Takuya \textsc{Akahori},\altaffilmark{3}
             Motokazu \textsc{Takizawa},\altaffilmark{4}
             Yutaka \textsc{Fujita},\altaffilmark{5}
             and Kazuo \textsc{Makishima}\altaffilmark{1,6,7}  }
\altaffiltext{1}{Departure of Physics, The University of Tokyo, 7-3-1 Hongo, Bunkyo-ku, Tokyo, 113-0033 Japan}
\altaffiltext{2}{SRON Netherlands Institute for Space Research, Sorbonnelaan 2, 3584 CA Utrecht, the Netherlands}
\altaffiltext{3}{Kagoshima University, 1-21-35, Korimoto, Kagoshima 890-0065, Japan}
\altaffiltext{4}{Department of Physics, Yamagata University, Kojirakawa-machi 1-4-12, Yamagata 990-8560, Japan}
\altaffiltext{5}{Department of Earth and Space Science, Graduate School of Science, Osaka
University, Toyonaka, Osaka 560-0043, Japan}
\altaffiltext{6}{Research center for the Early Universe, The University of Tokyo, 7-3-1  Hongo, Bunkyo-ku, Tokyo, 113-0033, Japan}   
\altaffiltext{7}{MAXI Team, RIKEN, 2-1 Hirosawa, Wako, Saitama 351-0198, Japan}

\maketitle

\begin{abstract}
CIZA J1358.9-4750 is a nearby ($z=0.074$) pair of clusters of galaxies located close to  the Galactic plane. It consists of two X-Ray extended humps at north-west and south-east separated by 14 arcmin  ($\sim 1.2$~Mpc), and an X-Ray bright bridge--like structure in between. With Suzaku, the south-east hump was shown to have a temperature of $5.6\pm0.2$~keV and the north-west one $4.6\pm0.2$~keV. Neither humps exhibit significant central cool component. The bridge region has a temperature higher than $9$~keV at the maximum, and this hot region is distributed almost orthogonal to the bridge axis in agreement with the shock heating seen in numerical simulations at an early phase of a head-on major merger. This resemblance is supported by good positional coincidence between the X-Ray peaks and cD galaxies associated with each cluster. In a short exposure XMM--Newton image, a significant intensity jump was found at a position where the Suzaku-measured temperature exhibits a steep gradient. These properties indicate the presence of a shock discontinuity. The Mach number is estimated to be $1.32\pm0.22$ from the temperature difference across the identified shock front, which gives the colliding velocity of approximately $1800$~km s$^{-1}$. From optical redshifts of the member galaxies, the two clusters are indicated to be merging nearly on the sky plane. Thus, CIZA J1358.9-4750 is considered as a valuable nearby example of early-phase merger with a clear shock feature.
\end{abstract}

\section{Introdcution}
Galaxy clusters are considered to grow through their mergers. 
Huge gravitational energy up to $\sim 10^{64}$ erg is released during the event,
especially in case of ``major merger'', i.e, nearly head-on merging of two clusters with similar sizes (e.g. Ricker and Sarazin 2001).
Part of this energy is converted into thermal energy of the intracluster medium (ICM), mainly via collision-induced shocks. These shocks also convert the energy into non-thermal energy such as those to accelerate cosmic-rays, drive ICM turbulence, and amplify intracluster magnetic fields. However, it is still unknown how the released dynamical energy is distributed into these different channels.

One of the approaches to tackle this problem is to observe merging clusters with clearly identified collision parameters, such as colliding velocity, viewing angle, and the original mass of the colliding clusters. Numerical simulations suggest that shocks with the Mach number of $M=$ 3--5 arise in late phases (i.e. after core crossing) of cluster mergers (e.g. Ricker and Sarazin 2001). Such shocks, characterized by significant density/temperature jumps and sometimes with radio emissions (e.g. Feretti 2012), have been studied intensively with X-Ray and radio observations. It is, however, not easy to determine the collision parameters in late phase mergers, since 
original ICM temperature and density before shocks passage is not clear.
  An example is the nearby late phase merging clusters Abell 3667; it exhibits clear evidence for shock heating of the ICM (e.g. Briel et al. 2004, Nakazawa et al. 2009, Finoguenov et al. 2010, Akamatsu et al. 2013) together with two bright diffuse radio emission, namely relics (e.g. Roettiger et al 1999), but its merger geometry is still unclear. 
In case of another famous late phase merger, the Bullet cluster (1E0657-56), the merger geometry looks clear (e.g. Markevitch et al. 2002) although the original ICM temperature itself is not well determined.

We can more easily understand the merger geometry in early phase (i.e before core crossing) merging clusters, such as Abell 222-223 (e.g. Werner et al. 2008), Abell 399-401 (e.g. Fujita et al. 1996, Fujita et al. 2008) pairs 
although they do not show strong evidence for shock heating yet.
Cygnus-A cluster of galaxies, for example, is one of a few pre-core crossing nearby merger with evidence for shock heating in between the two cores (Markevitch et al. 1999, Sarazin et al. 2013). In this case, there is no clear detection of surface brightness jump, and the shock location is rather ambigious.


In the present work, we report on Suzaku and XMM-Newton observations of an early phase merger CIZA J1358.9-4750 which clearly exhibits shock signature. We use the Hubble constant $H_0$ = 67 km~s$^{-1}$~Mpc$^{-1}$, together with the cosmological parameters $\Omega_M=0.31$ and  $\Omega_\Lambda=0.69$ (Planck Collaboration, Ade, P. A. R., Aghanim, N., et al. 2014). 
The errors refer to $90\%$ confidence for a single parameter.


\section{CIZA J1358.9-4750}
This object, CIZA J1358.9-4750, is listed in the CIZA catalog (Ebeling et al. 2002, Kocevski et al. 2007), which collects X-Ray selected clusters of galaxies in the Zone of Avoidance (typically ${\mid}$b${\mid}$ $<$ $20$°, where b is the Galactic latitude).  By inspecting ROSAT and XMM--Newton short exposure images, $3.2$~ks and $4.8$~ks respectively, we identified this object as a nearby (redshift of $\sim0.07$, that is,  angular diameter distance of $290$~Mpc) major merger candidate.
 The $0.1$--$2.4$ keV ROSAT flux reaches $1.92\times10^{-11}$~ergs cm$^{-2}$ s$^{-1}$   which is the sixth highest in the CIZA catalog excluding point source candidates.
It has two distinct X-Ray humps in the north-west (NW) and south-east (SE) directions, separated by $\sim14$~arcmin (approximately $1.2$~Mpc), which are both considered to represent clusters of galaxies. Furthermore, a clear X-Ray enhancement is seen to connect them.
A bright elliptical galaxy, considered to be the cD galaxy, is located right at the center (within $1$~arcmin~=~$84$~kpc) of each X-Ray hump. The two galaxies (2MASX J13590381-4751311 and 2MASX J13581085-4741243) at the center of the SE and NW clusters have optical redshifts of $0.0745$ and $0.0709$, respectively (from NED database).

\section{SUZAKU Observation}
\begin{figure}
 	\begin{center}
		\FigureFile(80mm,50mm){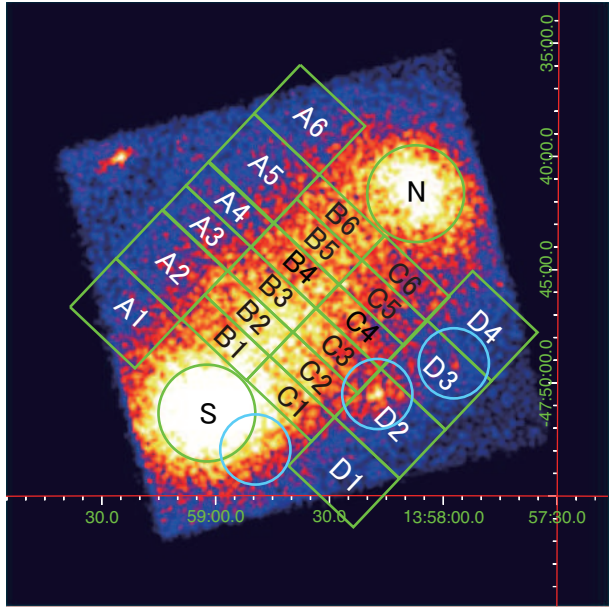}
	 \end{center}
	 \caption{A $0.6$--$10$~keV X-Ray image of CIZA J1358.9-4750 taken with Suzaku XIS3, shown without background subtraction.  Spectra are extracted from regions indicated in green. Light-blue circles indicate the 3 point sources  discussed in section 3.1.} 
	 \label{image}
\end{figure}

With Suzaku, we observed CIZA J1358.9-4750 on 2013 January 21-23, covering the two X-Ray humps.
The XIS was operated in the normal full-frame clocking mode. 
In this paper, we employ the data of both 3$\times$3 and 5$\times$5 edit modes acquired with XIS0 and XIS3.
With the standard data screening, we obtained an exposure of $61.7$~ksec for the XIS.
The $0.6$--$10$~keV image of the XIS3 detector is shown in figure \ref {image}.
It clearly reconfirms the two X-Ray humps and the bridge-like X-Ray emission between them.

\subsection{Spectral analysis}
As shown in figure \ref {image}, we defined 24 regions 
shown in green
for our spectral analysis. Among them, circles labeled N and S are for the NW and SE clusters, respectively, with a radius $2'.17$  ($180$~kpc) centered on the X-Ray peaks. The regions B1--B6 and C1--C6 cover the central parts (near the suggested collision axis) of the bridge, augmented by the series A1--A6 and D1--D4. Each region has a size of $4'.0 \times 1'.5$ or $3'.0$ (335~kpc $\times$ 130 or 260 kpc). Then, using XIS0 and XIS3, on-source spectra were accumulated over the 24 regions individually.
The XIS0 and XIS3 data were added together, and instrumental background (non X-Ray background: NXB) was estimated from night Earth data by xisnxbgen (Tawa et al. 2008). The background was then subtracted from the spectra of each region.
Note that because the regions A5 and A6 happen to be on the XIS0 segment-A, which is not well working, we only used XIS3 data.
 
We consider three sky background components; the local hot bubble (LHB) emission with a temperature of $0.1$~keV, that from the Milkyway halo (MWH) with $0.3$~keV, and the cosmic X-Ray background (CXB). 
The temperatures of LHB  and MHW were determined from an off-source Suzaku spectrum, obtained in an observation of the star HD125599, which is located 4$^\circ$ away from CIZA J1358.9-4750 at a similar Galactic latitude.
The normalizations of LHB and MHW were determined by fitting all the 24 extracted spectra of CIZA J1358.9-4750 simultaneously, assuming LHB and MWH have a constant surface brightnesses in the observed regions.
The $2$--$10$~keV surface brightness of the CXB was assumed as $6.38\times10^{-8}$~erg $s^{-1}$ cm$^{-2}$ sr$^{-1}$ using the parameters of Kushino et al. 2002, 
and its photon index was fixed at $\Gamma=1.41$.
All these parameters are listed in table \ref{backcomp}.

For spectral fits response matrix files and ancillrary response files were calculated using xisrmfgen and xissimarfgen, respectively (Ishisaki et al. 2007). 
By fixing the parameters of the background components as determined above, the 24 NXB-subtracted spectra in energy band of $0.6$--$10$~keV were fitted individually by a model of apec[LHB] + wabs $\times$ (apec[MHW] + powerlaw[CXB] + apec[ICM]) [model in XSPEC].
The interstellar absorption was represented by the wabs factor, where in hydrogen column density fixed at $N_H=11.5 \times 10^{20}$ cm$^{-2}$ as taken from the 21cm measurments (Kalberla,~P.~M.~W. et al. 2005). 
We left free the temperature, abundance, redshift, and norm parameters of the ICM, but the abundance and redshift parameters were fixed as 0.2 and 0.07, respectively, in low statistic spectra, i.e, having count rates less than $2.0\times10^{-2}$ cts s$^{-1}$. An energy band of $1.7$--$1.9$ keV was ignored because of the prominent instrumental feature. The spectra of S, N, A3 and A4, B3 and B4 are shown in Fig.\ref{spectral1}. The 24 fits were all successful, with $\chi^{2}$/d.o.f in the range of 0.79 to 1.18, and yielded the best-fit parameters as given in table \ref{parameters}. 
 Thus, the ICM temperatures of the individual regions have been determined with a typical $90\%$ error of $\pm1$~keV.
{\tabcolsep=4mm 
\renewcommand\arraystretch{1.2} 
\begin{table}[htb]
\caption{Summary of the fixed prameters for background components.}
	\begin{center}
 		 \begin{tabular}{lclclclcl} \hline 
					&	LHB		&	MHW	&	CXB\\ \hline \hline
$kT$(~keV)/$\Gamma$				&	0.10		&	0.29		&	(1.41)	\\
norm ($\times$10$^{-3}$)$^{\dagger}$ &	35.61	&	2.59		&	(0.96)\\ \hline

	\end{tabular}
  		\label{backcomp}
 		
  	\end{center}
	\begin{center}
	$^{\dagger}$ : nomalization of apec and powerlaw models in XSPEC ver12.8.0 scaled with factor $1/400\pi$.
	\end{center}
\end{table}

\begin{figure*}[htb]

	\begin{minipage}{0.33\hsize}
		\begin{center}
			\FigureFile(55mm,39mm){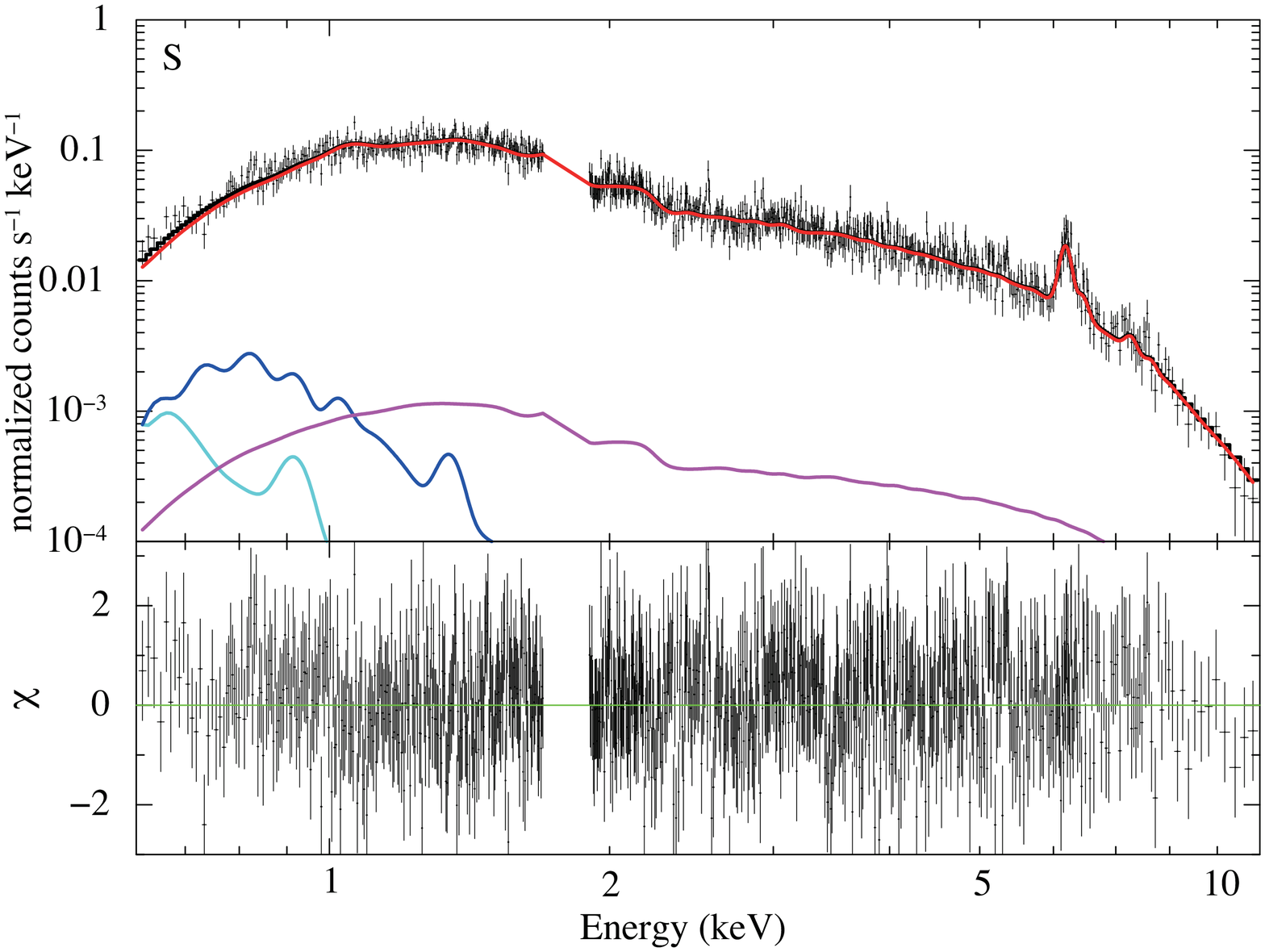}
	 	\end{center}
	\end{minipage}
	\begin{minipage}{0.33\hsize}
		\begin{center}
			\FigureFile(55mm,39mm){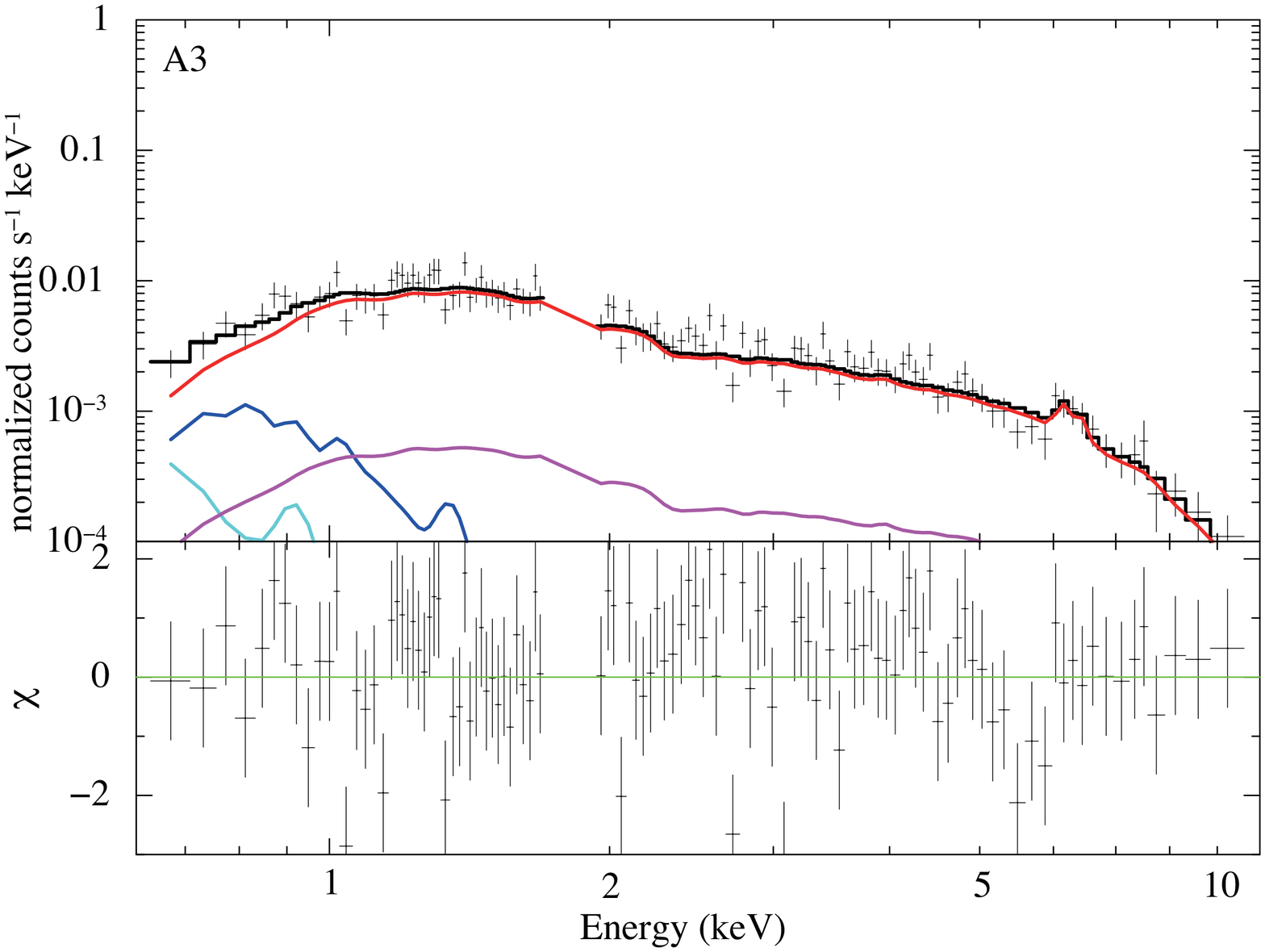}
	 	\end{center}
	\end{minipage}
	\begin{minipage}{0.33\hsize}
		\begin{center}
			\FigureFile(55mm,39mm){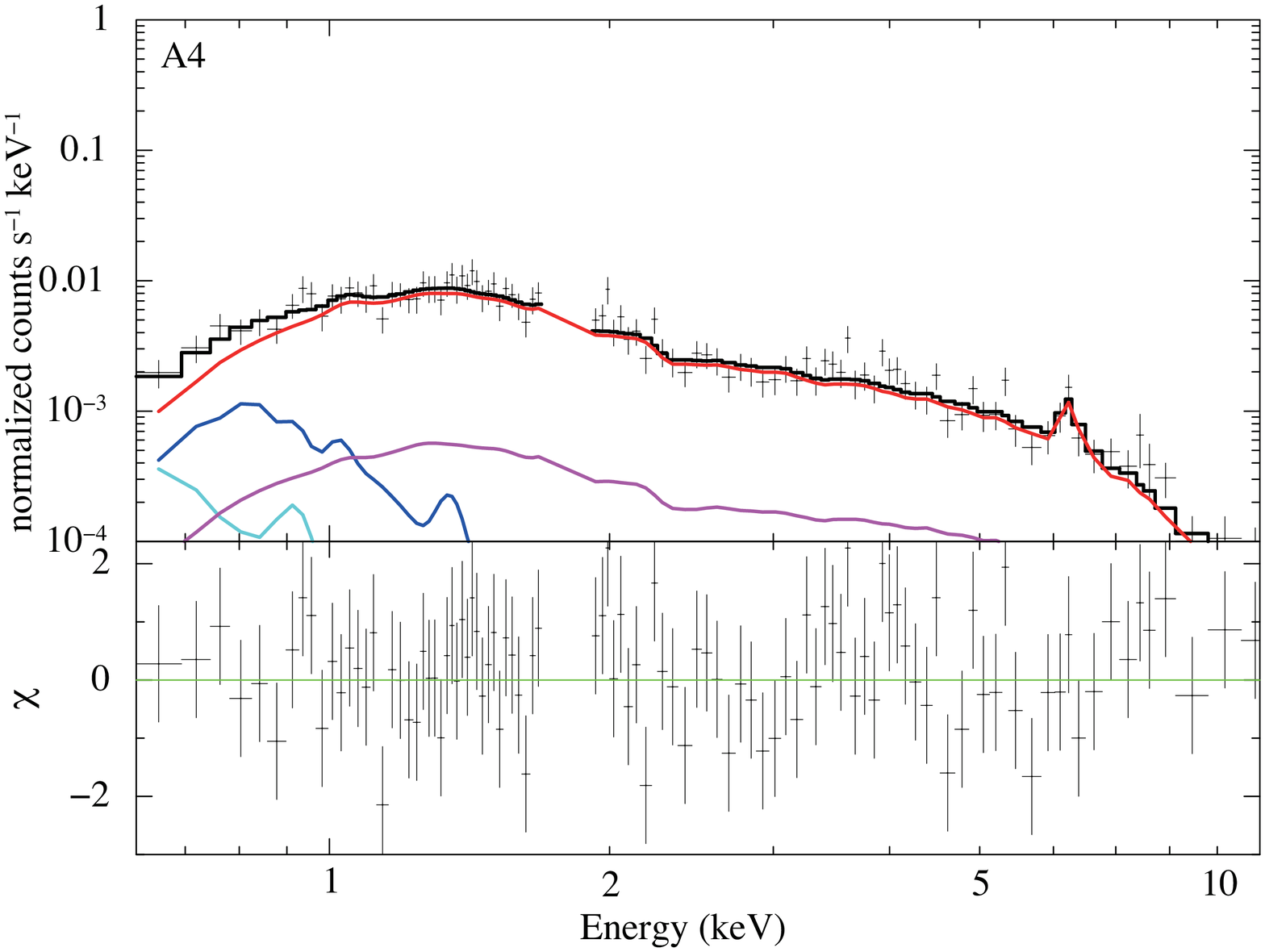}
	 	\end{center}
	\end{minipage}
	\begin{minipage}{0.33\hsize}
		\begin{center}
			\FigureFile(55mm,39mm){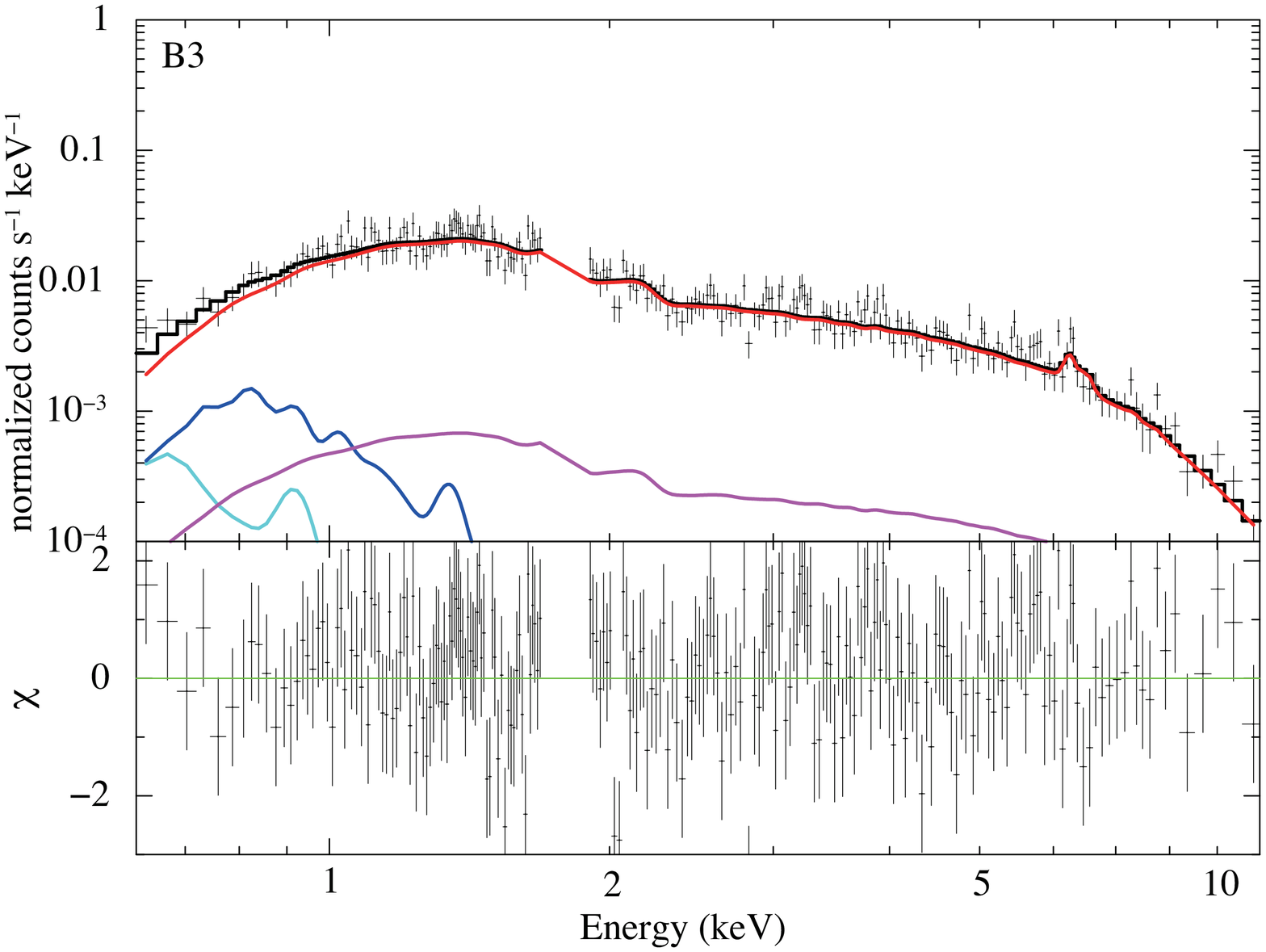}
	 	\end{center}
	\end{minipage}
	\begin{minipage}{0.33\hsize}
		\begin{center}
			\FigureFile(55mm,39mm){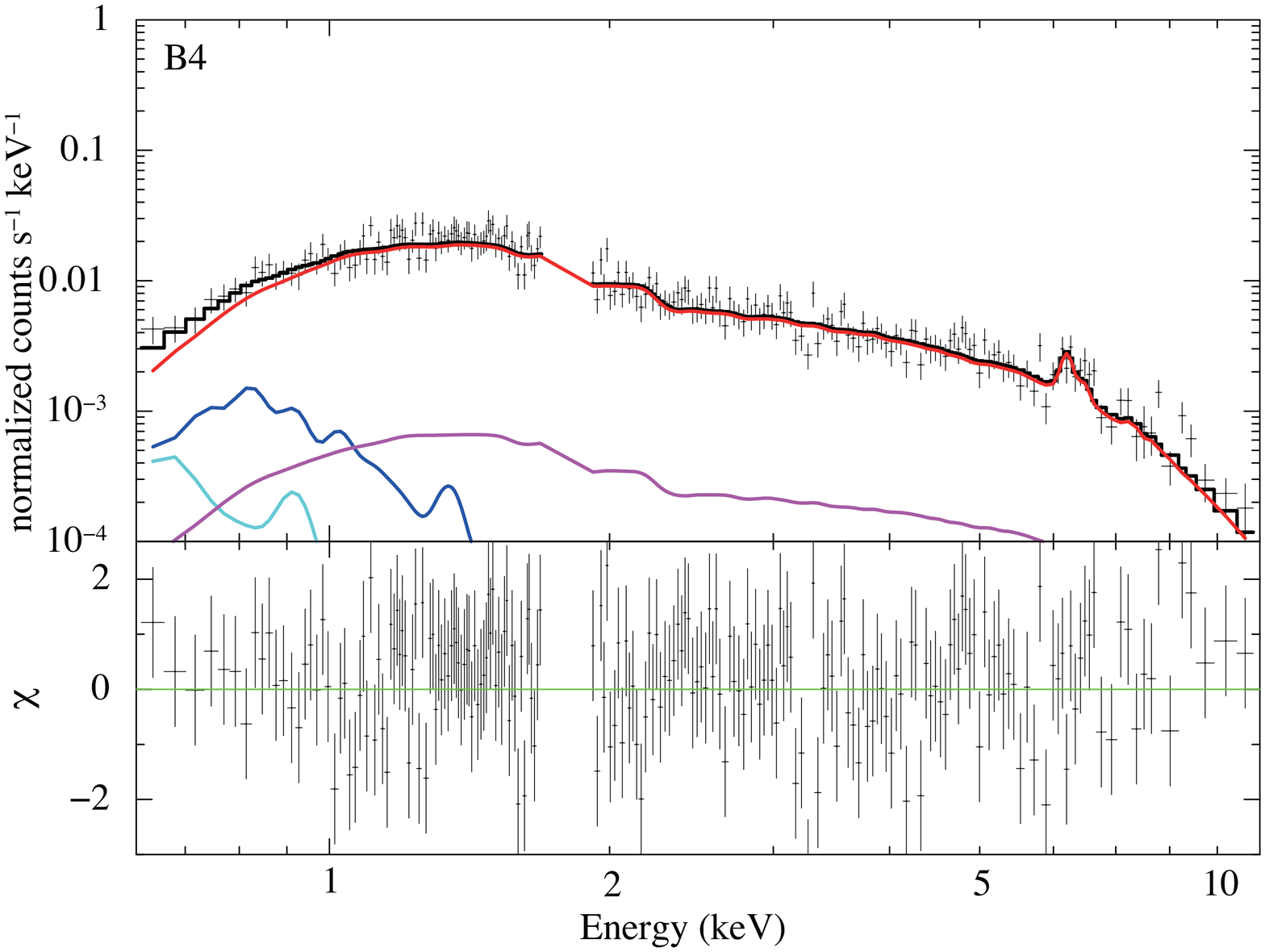}
	 	\end{center}
	\end{minipage}
	\begin{minipage}{0.33\hsize}
		\begin{center}
			\FigureFile(55mm,39mm){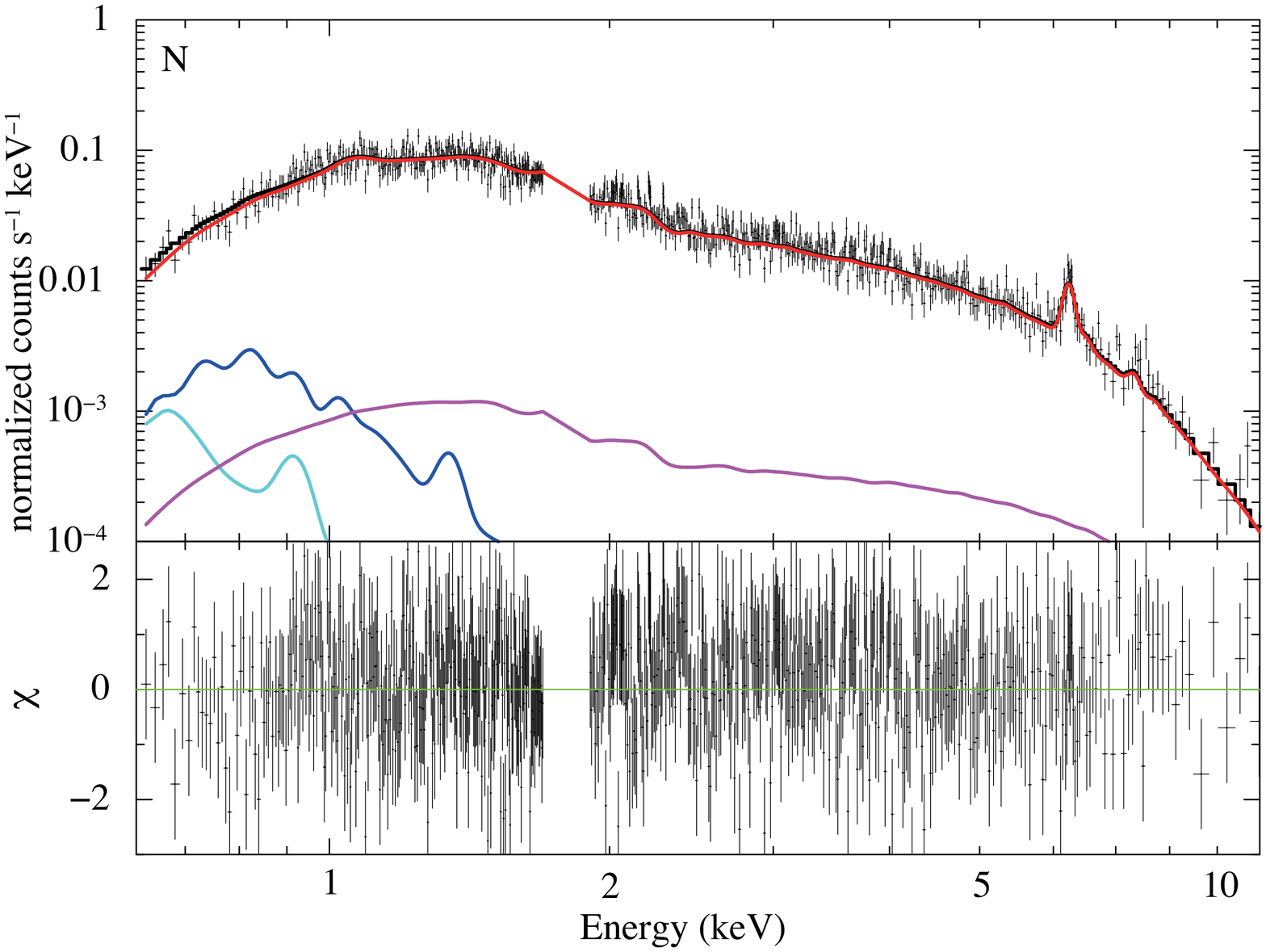}
	 	\end{center}
	\end{minipage}

  
\caption{Suzaku (XIS0+XIS3) spectra of CIZA J1358.9-4750, extracted from regions labeled S, N, A3, A4, B3, and B4 in figure. Red, cyan, dark blue and magenta indicate the ICM, LHB, MHW, and CXB components, respectively.}
\label{spectral1}
\end{figure*}

{\tabcolsep=4.5mm 
\renewcommand\arraystretch{1.3} 
\begin{table*}[htb]
	\caption{The parameters of Temperature, Abundance, normalization. The letters specify the regions shown in figure \ref{image}.}
	\begin{center}
	
 		 \begin{tabular}{c|cccccc} \hline  
			&		A1		       &		A2	        	     &			A3	      &		A4		     &			A5	         &		A6	\\	 \hline  \hline
$kT$(~keV)     	&$6.50_{-0.72}^{+0.93}$&$7.51_{-0.84}^{+0.92}$&$9.26_{-1.31}^{+2.07}$&$6.56_{-0.88}^{+1.32}$&$8.22_{-1.30}^{+1.97}$&$5.70_{-0.88}^{+1.06}$	\\
Z$\odot$ 		&$0.38_{-0.17}^{+0.18}$&$0.22_{-0.11}^{+0.16}$&fixed&fixed&$0.28_{-0.20}^{+0.26}$&fixed	\\
$\chi^{2}$/d.o.f &$137/147$&$212/205$	&$130/112$&$88/100$&$108/115$	&$69/77$	\\ \hline \hline
		
			&	B1		&	B2	&	B3	&	B4	&	B5	&	B6	\\	 \hline  \hline			
$kT$(~keV)      	&$7.31_{-0.59}^{+0.74}$&$8.36_{-0.68}^{+0.87}$&$9.14_{-1.22}^{+1.17}$&$6.97_{-0.61}^{+0.75}$&$7.17_{-0.70}^{+0.79}$&$6.37_{-0.54}^{+0.58}$\\
Z$\odot$ 		&$0.36_{-0.10}^{+0.12}$&$0.46_{-0.14}^{+0.15}$&$0.25_{-0.12}^{+0.14}$&$0.29_{-0.10}^{+0.13}$&$0.19_{-0.10}^{+0.12}$&$0.18_{-0.09}^{+0.10}$	\\
$\chi^{2}$/d.o.f	&$267/248$&$235/249$&$244/233$&$222/210$&$218/211$&$183/217$\\  \hline \hline

			&	C1		&	C2	&	C3	&	C4	&	C5	&	C6	 \\	 \hline  \hline											
$kT$(~keV)      	&$6.29_{-0.46}^{+0.49}$&$6.81_{-0.55}^{+0.74}$&$8.20_{-0.81}^{+1.13}$&$6.66_{-0.61}^{+0.83}$&$6.60_{-0.62}^{+0.83}$&$6.11_{-0.99}^{+1.99}$	\\
Z$\odot$  		&$0.30_{-0.10}^{+0.10}$&$0.20_{-0.09}^{+0.10}$&$0.27_{-0.12}^{+0.13}$&$0.18_{-0.11}^{+0.13}$&$0.25_{-0.11}^{+0.13}$&$0.11_{-0.09}^{+0.12}$\\
$\chi^{2}$/d.o.f	&$262/245$&$207/224$&$192/205$&$178/186$&$160/172$&$167/165$	\\  \hline \hline

			&	D1		&	D2	&	D3	&	D4	&	S	&	N	 \\	 \hline  \hline	
$kT$(~keV)      	&$6.55_{-0.87}^{+1.07}$&$5.06_{-0.79}^{+1.02}$&$6.26_{-0.76}^{+1.59}$&$5.89_{-0.74}^{+0.83}$&$5.60_{-0.19}^{+0.19}$&$4.55_{-0.16}^{+0.16}$\\
Z$\odot$  		&fixed&fixed&$0.19_{-0.09}^{+0.16}$&$0.21_{-0.14}^{+0.16}$&$0.39_{-0.02}^{+0.06}$&$0.25_{-0.04}^{+0.05}$	\\
$\chi^{2}$/d.o.f	&$89/112$&$70/72$&$159/160$&$126/131$&$795/800$&$633/629$	\\ \hline

 		\end{tabular}

  		\label{parameters}

  	\end{center}
\end{table*}

}

There are 3 point sources with the $0.4$--$7.2$~keV X-Ray flux higher than $1.0\times10^{-14}$~ergs~cm$^{-2}$~s$^{-1}$,  estimated from the XMM-Newton short exposure data, as detailed in section 4. By excluding regions $1'.5$ around these sources, we estimated their contamination effects to our Suzaku temperature analysis. Only the result from region D2 changed significantly ($\sim1$~keV), while the effect in other regions is negligible, i.e. less than $0.1$~keV. Because we have cut the 3 point sources, the CXB normalization could be smaller than those provided by Kushino et al. 2002. Thus, we also checked if the results can change with the smaller CXB normalization derived from Lockman hole observation with almost no source contamination ($5.7\times10^{-8}$~erg $s^{-1}$ cm$^{-2}$ sr$^{-1}$, Suzaku observation ID 101002010; see Nakazawa et al. 2009) and found it to be insignificant, i.e. less than 0.1 keV. From the above, we replaced the results for region D2 using the source-excluded spectra. We also checked the effect of fixing {\it N$_H$} through-out this analysis, and also found that the difference cause by making the value free is negligible (less than $\sim 0.4$~keV) to the spectral parameters of the ICM component.


\subsection{Temperature distribtuion}
The derived temperature distribution is plotted in figure \ref{temp}. Center of the SE and NW clusters were found to have a temperature of $5.6\pm0.2 $~keV and $4.6\pm0.2$~keV, respectively. The small difference, approximately 20\%, indicates that they have similar mass from {\it M -- T} relation (Vikhlinin et al. 2006). These temperature suggest that Virial radius, i.e. $r_{200}$, of the SE and NW clusters are $1.7$~Mpc and $1.5$~Mpc, respectively. Using $\beta$-model assuming $\beta = 0.6$ and $r_{\rm c} = 180$~kpc at $0.5$--$10$~keV, luminosity within $r_{200}$ of the former cluster is estimated to be $2.6\times10^{44}$~erg s$^{-1}$, scaled from the central 180 kpc flux. Similarly, those of the NW cluster is estimated to be $1.8\times 10^{44}$~erg s$^{-1}$. To search for a signature of a cooling core, we extracted spectrum of N and S within a radius of $90$~kpc ($1'.08$), and derived temperatures of $5.7\pm0.3 $ and $4.4\pm0.3$, respectively. Therefore, neither of them show strong evidence for a cool core.

Compared to the two clusters, A3, B3 regions in the bridge region show higher temperatures typically by a factor of $\sim$1.5. The hot region extends to almost perpendicular to the axis connecting the two cores, namely, A3, B3, and C3 regions. The westward regions D1--D4 appear relatively isothermal. The measured temperatures exhibit a particularly significant increase from $\sim7$~keV in the region A4--B4--C4 to $\sim9$~keV in A3--B3--C3. 


\begin{figure*}[htb]
	\begin{minipage}{0.49\hsize}
		\begin{center}
			\FigureFile(92mm,68mm){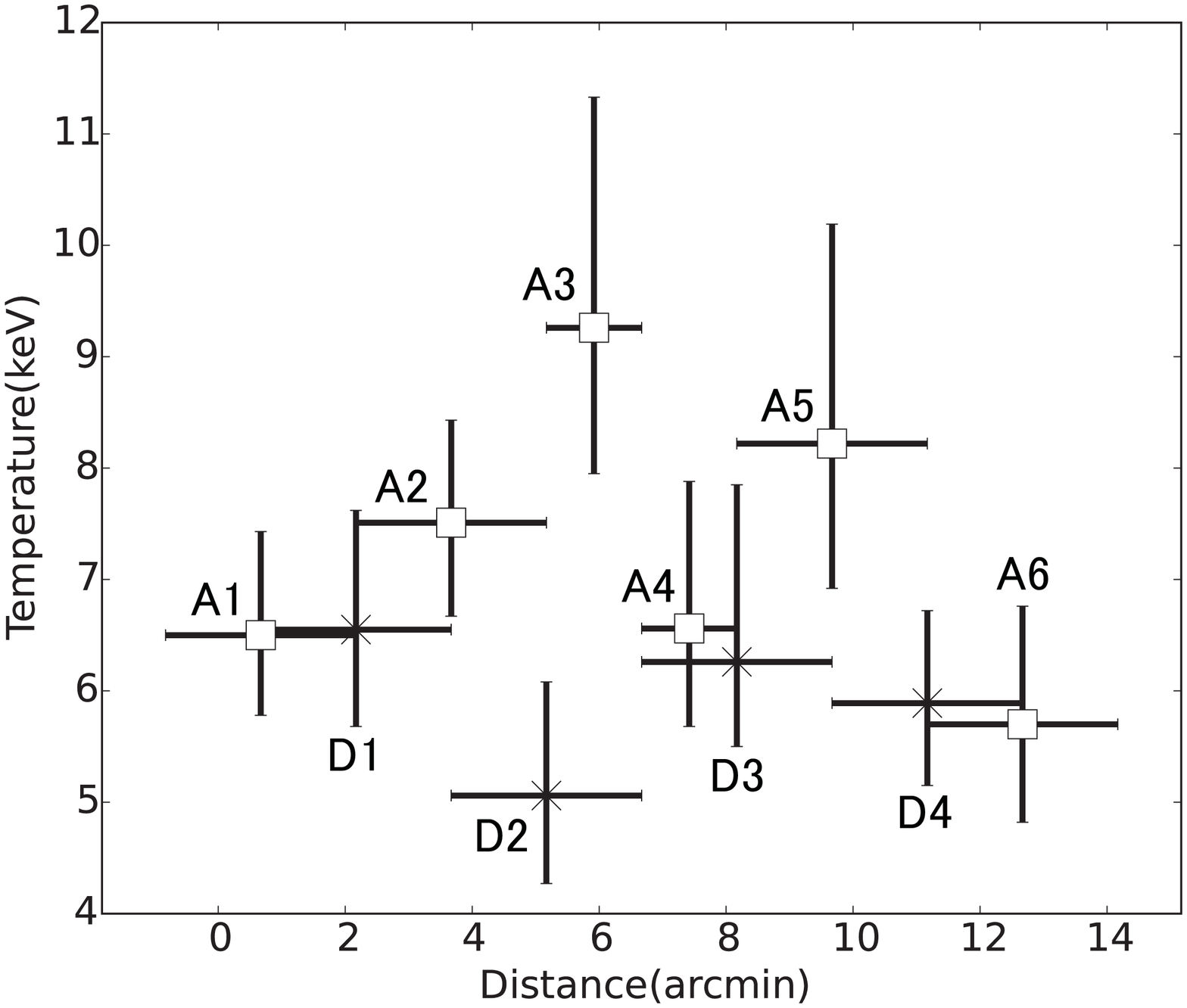}
	 	\end{center}
	\end{minipage}
	\begin{minipage}{0.49\hsize}
		\vspace{-2mm}
		\begin{center}
			\FigureFile(92mm,68mm){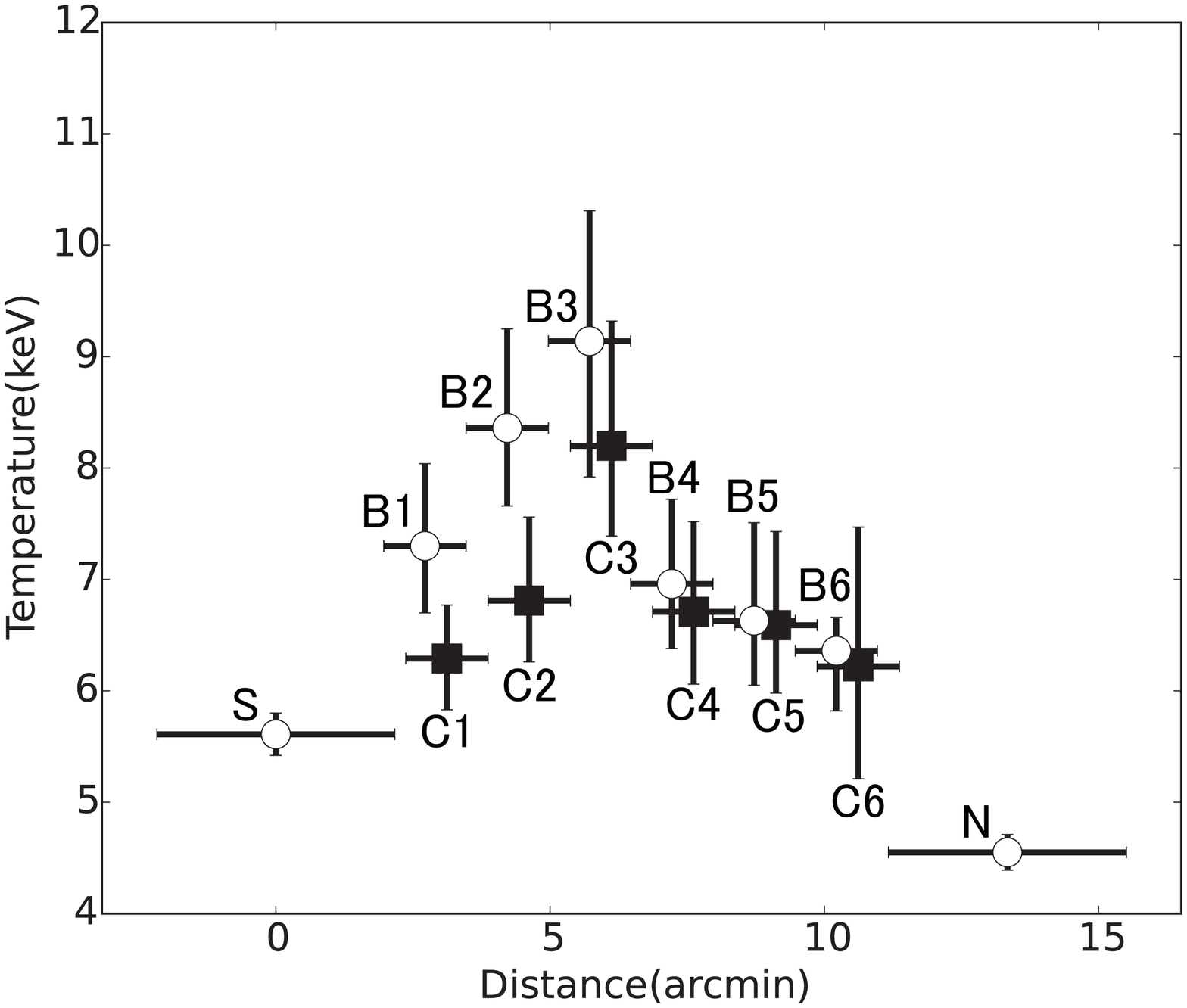} 
	 	\end{center}
	\end{minipage}
 	\caption{Temperature distributions of CIZA J1358.9-4750 in the A1--A6 and D1--D6 regions (left), and B1--B6 and C1--C6 (right). Abscissa represents the angular distance along the bridge axis. Note that statistics of the A5 and A6 regions are about a half, since we could not add data from XIS0.}
	\label{temp}
\end{figure*}

\subsection{Entropy distribution}
Sharp temperature increase is suggestive of the presence of a shock discontinuity. To identify whether this is caused by shock or adiabatic compression, entropy distribution are plotted in figure \ref{entropy}. For simplicity, we employ the ``astrophysical entropy'' expressed as $K=kT/n_e^{2/3}$.
Since it preserves heat input to the ICM, entropy is a good indicator of shocks. Here, we simply estimated the gas density by assuming a cylindrical geometry with a radius of $335$~kpc for regions B and C which is the width of these regions.  In regions A and D, more simpler cuboid geometry with line-of-sight depth of $670$~kpc and horizontal width of $335$~kpc $\times$ 130 or 250~kpc is adopted, 
because the over-all shape of the cluster around the periphery is not resolved. At regions of A4--B4--C4 to A3--B3--C3, the same place where temperature increase rapidly, the entropy shows a clear jump of a factor of $\sim1.4$. Thus, temperature increase is not only produced by adiabatic compression but also shock. There are few example of early merging clusters with clear evidence for shock. In addition, we also note that the shock is detected in relatively bright region and is easy to observe in detail thanks to its brightness. Therefore, this object is important to understand evolution of shocks as well as nature of shock.

\begin{figure*}[thb!]
	\begin{minipage}{0.49\hsize}
		\begin{center}
			\FigureFile(98m,70mm){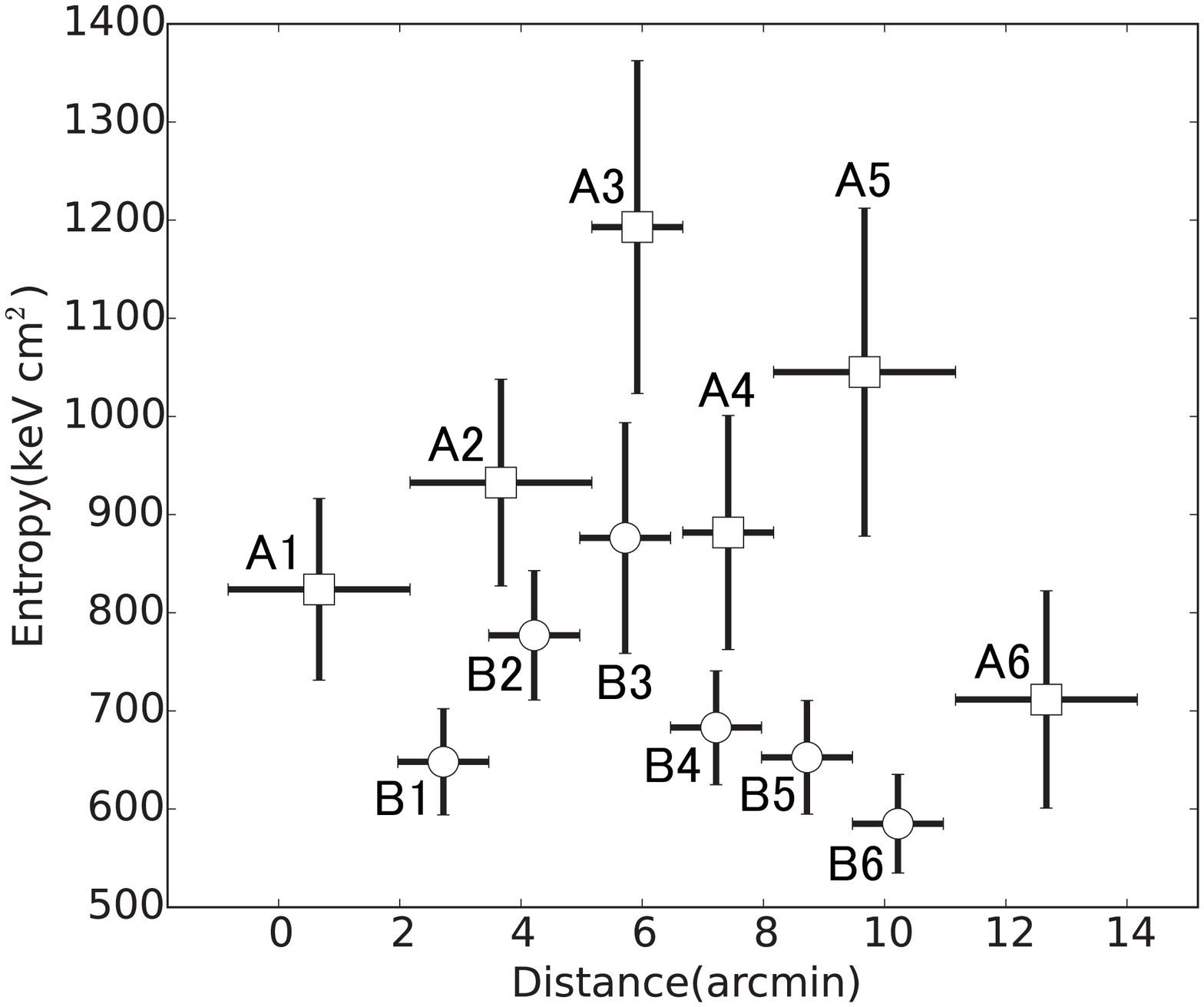}
	 	\end{center}
	\end{minipage}
	\begin{minipage}{0.49\hsize}
		\vspace{-2mm}
		\begin{center}
			\FigureFile(90mm,80mm){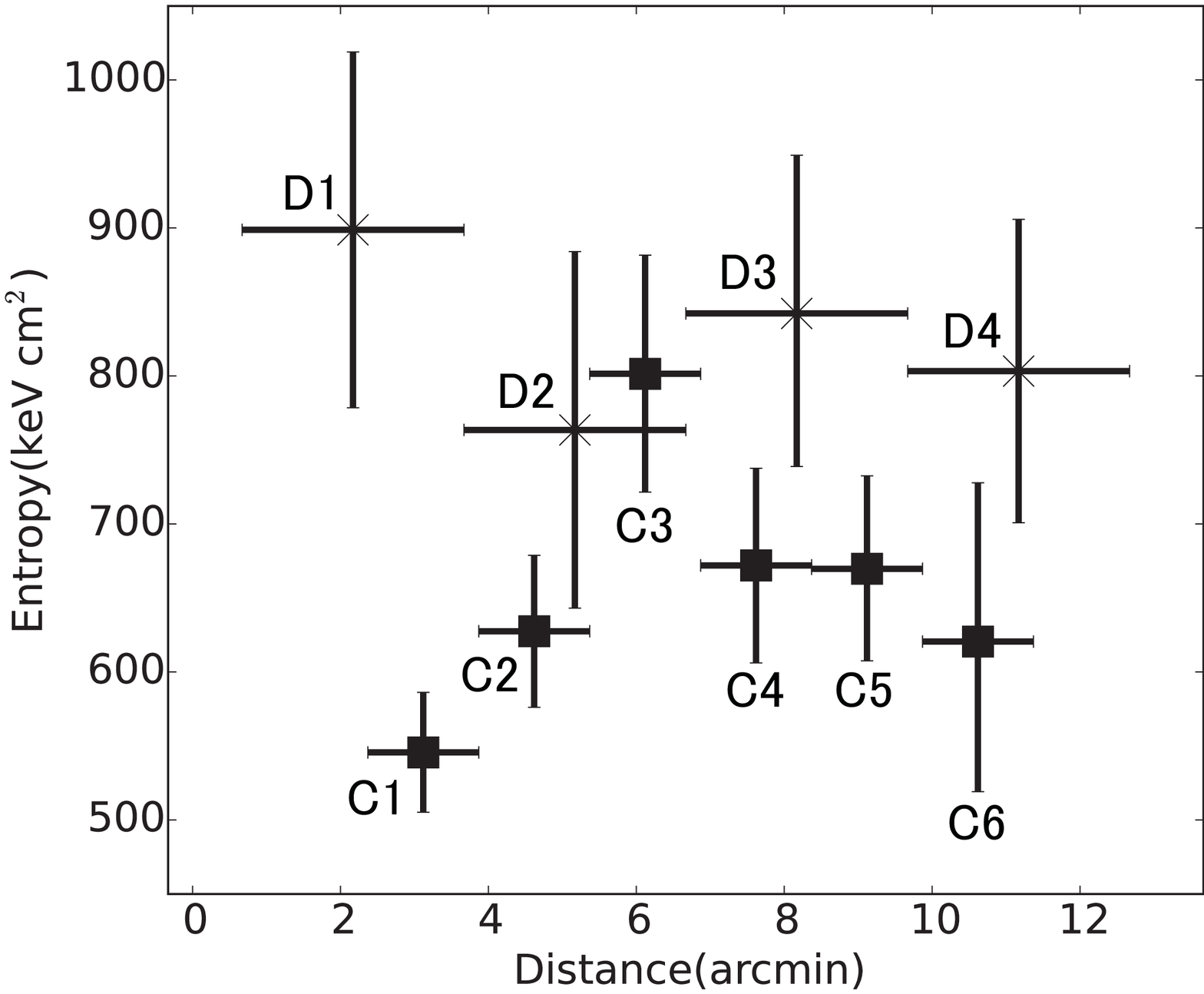}
	 	\end{center}
	\end{minipage}
 	\caption{Entropy distributions of CIZA J1358.9-4750 in the A1--A6 and B1--B6 regions (left), and C1--C6 and D1--D6 (right). Abscissa represents the angular distance along the bridge axis.}
	\label{entropy}
\end{figure*}

\section{Flux jump in the XMM-Newton image}
Shock is found from Suzaku entropy distribution, but further information about its detailed orientation is hampered by the limited angular resolution.
To search for a possible surface brightness jump caused by the shock suggested by figure \ref{entropy}, we further utilized an archival XMM-Newton snapshot data 
(ObsID = 0204710501; acquired in 2004 February 21). Basic reduction and calibration of the European Photon Imaging Camera (EPIC) 
data were carried out with SAS (ver. 13.5). In the screening process we set {\it FLAG} = 0, and kept events with {\it PATTERNs} 0--12 for 
the MOS cameras and those with {\it PATTERNs} 0--4 for the PN camera. By examining lightcurves extracted in $10.0-14.0$ keV and 
$1.0-5.0$ keV from source free regions, we discarded time intervals affected by hard-band and soft-band flares, respectively. The obtained MOS 
and PN exposures are 4.1 ks and 1.8 ks, respectively.

The raw event image in the $0.5$--$4.5$~keV band is shown in Fig.\ref{XMM}.  
We excluded 4 point sources selected with the same criteria as we did for the Suzaku spectral analysis.
 We then rotated the image by 25 degree counter-clockwise and produced brightness profiles to find a clear jump at the position $1500$~arcsec in the figure.  East region has biggest intensity jump and west is weaker. 
Although the XMM-Newton exposure is too short to determine the ICM temperature, the location of the brightness jump is right within the Suzaku high-temperature regions.

\begin{figure*}[!t]
	\begin{minipage}{0.49\hsize}
 		\begin{center}
			\FigureFile(80mm,80mm){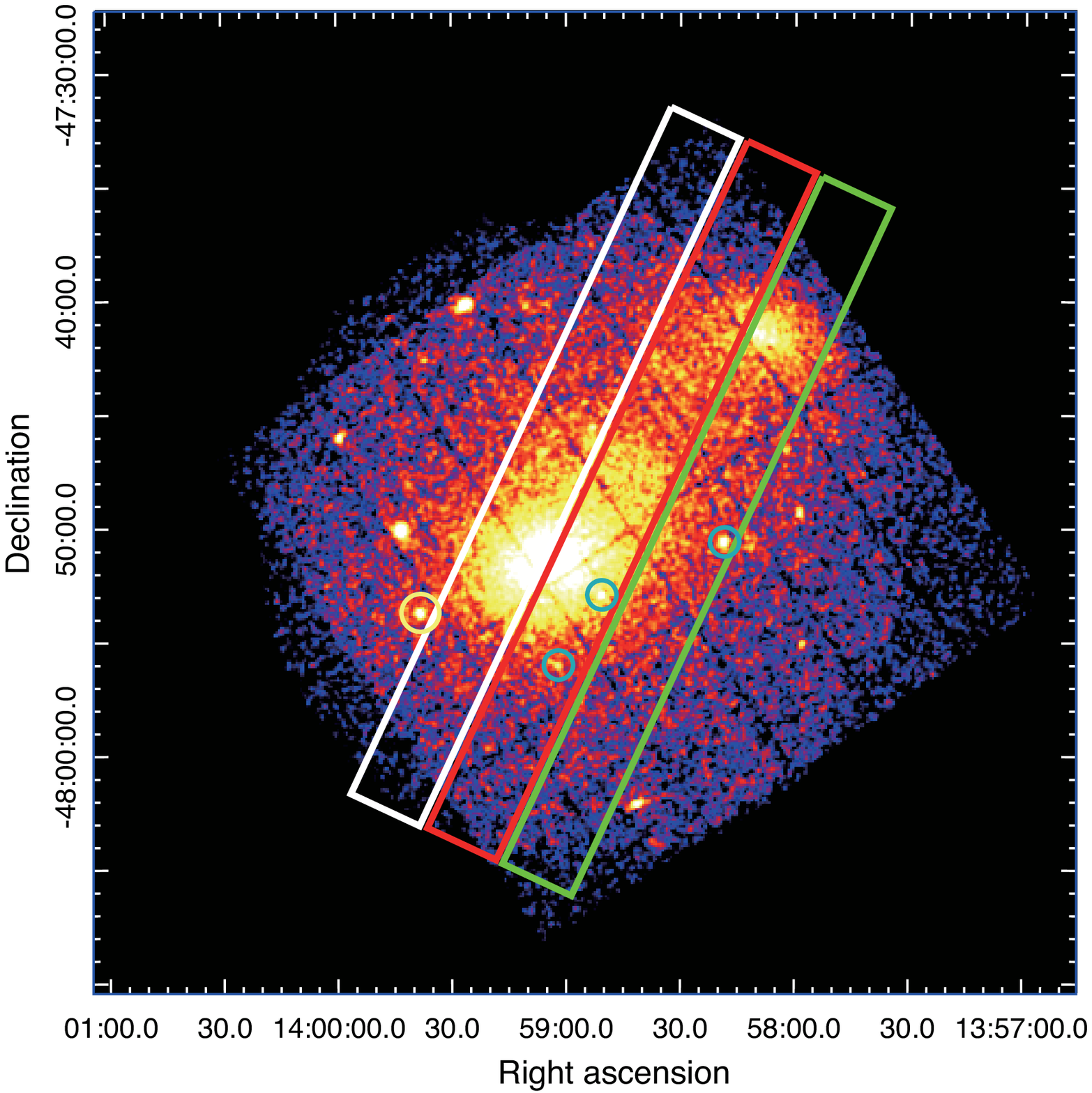}
		 \end{center}
	\end{minipage}
	\begin{minipage}{0.49\hsize}
 		\begin{center}
			\FigureFile(80mm,80mm){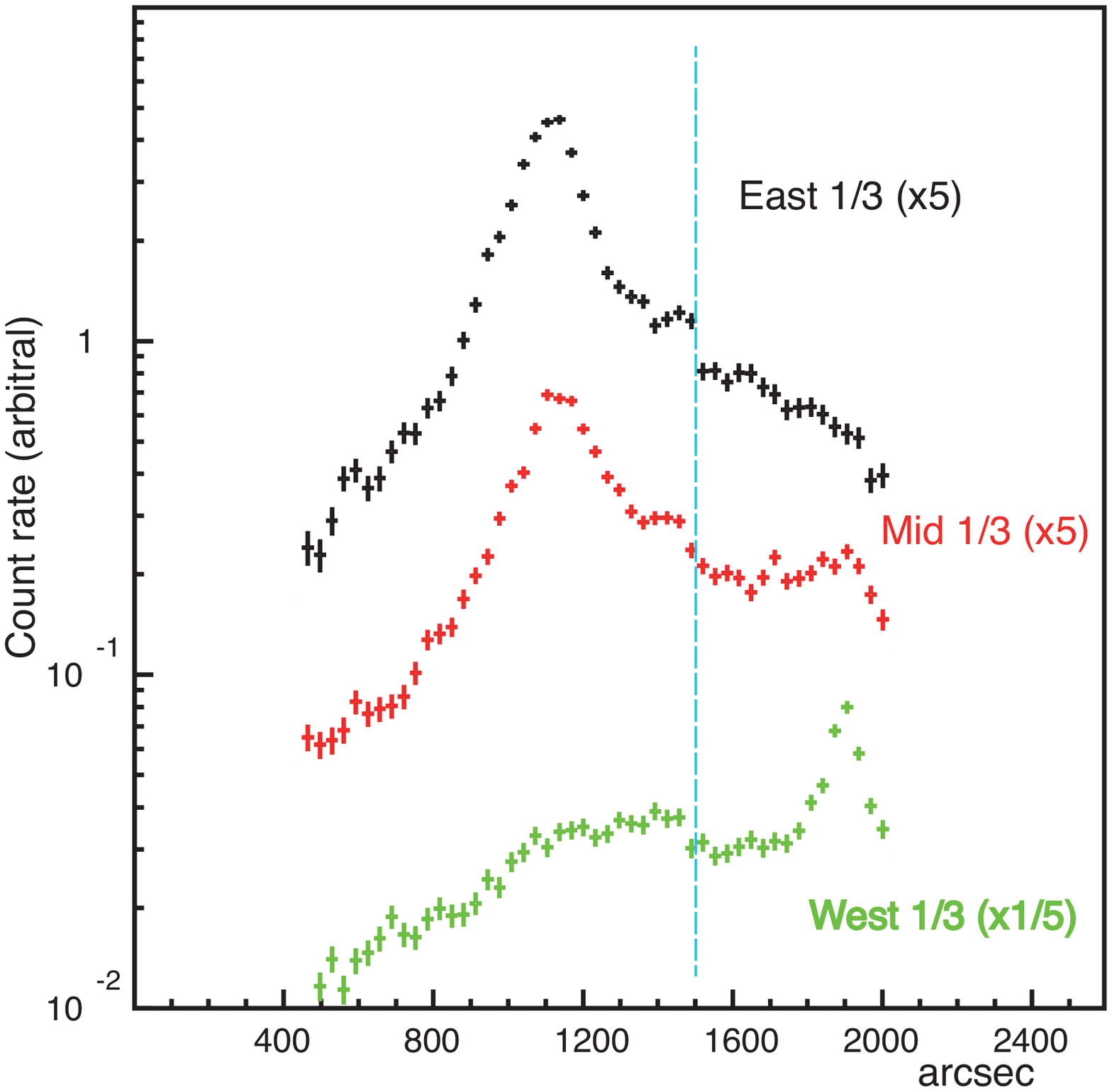}
 		\end{center}
	\end{minipage}
 	\caption{(a) A $0.5$--$4.5$~keV image of CIZA J1358.9-4750 with XMM-Newton. Surface brightness profiles are extracted from regions indicated in white, red and green, respectively. 
Small light--blue and yellow circles are the excluded regions around point sources with radius of $20^{\prime \prime}$ and $30^{\prime \prime}$, respectively. (b) $0.5$--$4.5$~keV surface brightness distributions of CIZA J1358.9-4750. Horizontal axis is in arcsec with arbitral zero point.}
	\label{XMM}
\end{figure*}


The shock front detected in the XMM-image is a bit inclined from the axis between the two peaks.
While the bridge axis is $43$~degree clockwise to the north, the shock normal is inclined by $25$~degree as noted above.
Because the region defined in Fig.\ref{XMM}(a) is aligned to the former, we also checked the temperature 
profile with a slightly rotated regions following the later angle, to find virtually the same results;
from SE to NW, it showed a temperature jump from $8.5_{-0.79}^{+1.31}$~keV to $7.5_{-0.85}^{+0.88}$~keV in regions corresponding to ``mid''. Note that both regions have a size of $1'.5\times3'.0$.

\section{Discussion}
\subsection{Merger phase}
Using Suzaku, we confirmed that the X-Ray image of CIZA J1358.9-4750 has two humps, corresponding to the two clusters without cool cores, and a bridge structure. In addition, we found that there is a temperature jump within the bridge. This X-Ray image and temperature distribution are similar to those seen in numerical simulations of an early phase of cluster major merger (e.g Akahori \& Yoshikawa 2010; Takizawa 2008). The good positional coincidence between the cD galaxies and the X-Ray peaks provides another support to this view, where in the late stage of merger the coincidence tends to be not observed (e.g., Abell 2163; Okabe et al. 2011).

As shown in Fig.\ref{XMM}(b), the X-Ray surface brightness profiles derived with XMM-Newton exhibit a clear jump at the position where the ICM temperature measured with Suzaku exhibits a steep change. This good positional coincidence  implies a pressure jump in that region, that is, the presence of a shock discontinuity. 
Thus we can apply the Rankine-Hugoniot  condition to this jump.
Employing the largest temperature increase, 
$6.97_{-0.61}^{+0.75}$~keV in B4 to $9.14_{-1.22}^{+1.17}$~keV in B3, we obtain a Mach number of $1.32\pm0.22$. Since the sound velocity in the ICM is $\sim1360~$kms$^{-1}$ at 6.97 keV, i.e, at the pre-shock region, the colliding velocity is approximately $\sim$1800~kms$^{-1}$. 
Because the width of the spectral fitting regions are as narrow as $1'.5$ which is smaller than the Suzaku  PSF of $2'$ half power diameter, the spectral variation shall be smeared. The true temperature jump therefore could be a bit higher. In other words the shock Mach number presented here is lower limit.

In the present study we fitted the ICM emission with only one APEC model (1~T), because it provided acceptable fits.  According to numerical simulations (e.g., Takizawa 2008; Akahori, Yoshikawa 2010), it is plausible that there are multi-temperature structures in the projection along the line of sight toward the shock-heated regions (A3-B3-C3), i.e. the shock-heated very hot component and the ambient, original component. Therefore, our assumption of single temperature may provide a lower limit on the temperature increase, and the Mach number and the colliding velocity could be higher than as quoted above.


Numerical simulation of merging clusters suggest that there should exist a forward-shock and reverse-shock as shown schematically in fig.5 of the review paper presented by Markevitch \& Vikhlinin (2007). In this observation, clear symptom of reverse-shock was not discovered. This could be merely by the lack of angular resolution (Suzaku), exposure (XMM-Newton) and strength of reverse-shock. But, in FIg.\ref{XMM}(b), east region exhibit brightness jump slightly at $120'' = 2'$~($170$~kpc) from the identified discontinuity, resembling the reverse shock. 
With Mach 1.32 and the sound velocity of 1360 km s$^{-1}$ in the ICM, post-shock velocity can be derived to be 1200 km s$^{-1}$. 
If we assume a 1:1 merger and the post shock region has no bulk motion as the simplest approximation,
the shock-front velocity on the sky plane becomes 1200 km s$^{-1}$ and the reverse shock has the same velocity from symmetry.
Then the narrow width of $170$~kpc suggests that the shock is young, i.e. only $\sim$700~Myr has passed since its birth.

 
  

\subsection{Merger geometry and its implication}
Because the shock front is clearly visible, the collision axis is considered to be nearly parallel to the sky plane. To confirm this, we utilized the NED database to identify the optical redshifts of the two clusters. We found $8$ and $11$ redshift-measured galaxies within $455$~kpc of the SE and NW clusters, respectively. The average redshifts are 0.0721 and 0.0737 in the NW and SW humps, respectively. Since the difference, $\Delta$ v = $480$ kms$^{-1}$, is significantly smaller than the estimated colliding velocity,  the two clusters are confirmed to be almost on the sky plane.

 One open question is the fact that the (candidate) cD galaxies have significantly different redshifts, with $\Delta t= 1000$~km s$^{-1}$. However, we do not have any additional information on this issue and leave it for future investigations.

\section{Summary and Conclusion}

We found that CIZA J1358.9-4750 is made of two medium sized clusters separated by $1.2$~Mpc which are connected by  a bridge of enhanced X-Ray emission. The bridge region is hotter than the main emission of each cluster and the position of temperature increase coincides with a brightness jump.
We identified this structure with a shock fronts, and derived a Mach number of $1.32\pm0.22$.
Thus, CIZA J1358.9-4750 is concluded to be a precious example of early-phase major merger with clear evidence of shock and shock heating, produced as the two clusters of similar sizes are making a head on collision along the sky plane.


\section*{AKNOWLEDGEMENTS}
This work was supported by JSPS KAKENHI Grant Number 26400218.



\end{document}